\documentclass{article}
\usepackage{microtype}

\usepackage{ijcai24}

\usepackage{times}
\usepackage{soul}
\usepackage{url}
\usepackage[hidelinks]{hyperref}
\usepackage[utf8]{inputenc}
\usepackage[small]{caption}
\usepackage{graphicx}
\usepackage{amsmath}
\usepackage{amsthm}
\usepackage{booktabs}
\usepackage{algorithm}
\usepackage{algorithmic}
\usepackage[switch]{lineno}
\usepackage{amssymb}

\title{Non cooperative Liquidity Games and their application to bond market trading}
\date{updated 9 Jan '24 }

\author{
Alicia Vidler$^1$
\and
Toby Walsh$^1$\and
\affiliations
$^1$UNSW, Australia\\
\emails
a.vidler@unsw.edu.au,
t.walsh@unsw.edu.au, \\
9th January 2024
}

\begin{document}
\maketitle

\begin{abstract}
    
We present a new type of game: the Liquidity Game. We draw inspiration from the UK government bond market and apply game theoretic approaches to its analysis. In Liquidity Games, market participants (agents) use non-cooperative games where the players' utility is directly defined by the liquidity of the game itself, offering a paradigm shift in our understanding of market dynamics.  Each player’s utility is intricately linked to the liquidity generated within the game, making the utility endogenous and dynamic. Players are not just passive recipients of utility based on external factors but active participants whose strategies and actions collectively shape and are shaped by the liquidity of the market. This reflexivity introduces a level of complexity and realism previously un-attained in conventional models.  We apply Liquidity Game theoretic approaches to a simple UK bond market interaction and present results for market design and strategic behaviour of participants. We tackle one of the largest issues within this mechanism - what strategy should market makers utilise when uncertain about the type of market maker they are interacting with and what structure might regulators wish to see. 
\end{abstract}



\section{The Liquidity Game: An introduction}
We introduce a game, the Liquidity Game (LG). We develop a game theoretic approach to analyse the opaque strategic interactions among UK bond market makers, focused on the transferal of bond holdings. We describe a new game formalism with endogenous utility functions, adding to this burgeoning class of games.  
The UK government bond market is mediated by 17 regulated market markets, know as Gilt-Edged Market Makers (GEMM's), through whom, in effect, all client transactions must flow.  For a variety of economic and regulatory reasons GEMM's cannot carry unlimited balances or holdings, (inventory) of government bonds. However, they are expected to continuously be available to clients to either buy or sell bonds, thus requiring GEMM's to have the financial capacity to transact with a client.  For reasons beyond the scope of this paper, such financial capacity means that GEMMS often need to transact between themselves when a accumulation or short fall of bonds at one GEMM becomes to great.  We describe and analyze the opaque interactions prevalent among market makers and propose a new type of strategic game (LG). The game is tailored to dissect the strategic transfer of asset (bond) holdings in an environment where market markers do not know against whom they play, nor the size of their opponent's holding. Understandably, players are prohibited from cooperation, and at no times may their asset balance change sign.  Players are compelled to base their actions on individual strategic considerations. Unlike many areas of game theory or finance, \textbf{price plays no part in this game}. This is relevant as Government bonds are priced in reference to prevailing interest rates and expectations, widely know information to market participants \cite{Duffie_1999}. Thus, the feature (utility derived by players) of the game is not the price of an asset, but rather the disposal or accumulation of a quantity of bonds, itself.  To quote the former Governor Warsh of the US Federal Reserve "Liquidity is confidence ". \footnote{https://www.federalreserve.gov/newsevents/speech/warsh20070305a.htm}


\subsection*{Liquidity Games:} 

\begin{itemize}
    \item We term market makers as agents, or players \(N_i\), and \(i = {1, 2}\) (by construction, we are modeling a bilateral mechanism).  Play is simultaneous.
    \item  No cooperation between players can occur.
\item The \(i\)th player holds an bond balance \(B_i \in \mathbb{Z}\),  such that \([x\leq B_i\leq y] , \forall i\) where  \(x,y \in\mathbb{Z}\).     

\item Quantities \(x\) and \(y\) are determined by the specific market and are known to all agents.  
\item The UK government makes public how many bonds are on issue, \(c\). Thus, \(min(x) = 0\) and \(max(y) = c\). 
\item Holdings, \(B_i\) are closely guarded private information and  \(U\) is an \(n\) -tuple of  \textbf{Utility functions}, which we define to be the value of units of bonds traded \(a_i\) such that \(u_i : \to \mathbb{Z}\) and where \(a_i\) is an action equal to a specific size of bond parcel played and  \([ 0 \leq a_i \leq |max(B_i)|] , \forall a_i\). 

\end{itemize}

    \begin{equation} \label{utilitySim}
        \text{Utility}(i ,B_i ) = |{a_i}| \quad  \forall \, a_i\in{A}
    \end{equation}

An \textbf{instance} of the game is the n-tuple \(I = (N, B , U) \)
\textit{We will interchangeable use the terms GEMM's, 'player' and 'agent' through out. Likewise, the term bond and asset will be used interchangeably.}

Permissible actions, \(a_i\) for each player is a set of asset quantities, choose strategically, defined as the size of bonds a player wishes to play, such that \(\forall i\), \({a_i} \in [0,|B_i|] \).  Actions, \(a_i\) are defined as the quantum of bonds that can be played by any agent such that \(a_i \in \mathbb{Z}, \forall a_i\).  Each agent, is incentivised to protect their true bond holding(\cite{Jerabeck2019}. Thus, they are likewise motivated to clear their balance in its entirety in one move.  Time is of the essence.


\subsection*{Summary of our contribution}
We focus on defining a new type of game, where the utility structure of this game diverges from conventional paradigms and specifically money, or asset price, play no part. We develop LG's whose utility function is determined by the efficacious reduction of asset holdings. Monetary exchanges and asset transfers at market prices are conspicuously absent, accentuating the core objective of asset holding reduction as a measure of utility and thus a definition of strategic action. In defining such, we contribute firstly generic game design, and secondly, an application of game theory to a novel domain, capital markets trading and design.

We explore the existence of an equilibrium, or solution concept for LG's.  Solution concepts in games applied to financial markets speak to the notion of stability. Stability is motivated here by trust of participants to continue to enter the market place and transact parcels of bonds as required - a crucial feature to a government bond market \cite{Pinter2023}.  A stable market is one where participants feel that trading will take a trusted amount of time.  The corollary being, as rational agents, they are "satisfied" with their engagement in the game.  Such satisfaction supports the notion that they will continue to engage, leading to a stable market dynamic.  This has no bearing on their profit or loss, or other financial aspect, but rather their continued participation.  Recall, for a GEMM it is a legal requirement of their regulation that they continue to participate in trading. 

The paper is structured as follows, section 2 begins with an overview of current game theory research, highlighting the unique contribution of this work.  Sections 3 discusses the generalised game rules.  Section 4 details game solution outcomes from application of a simple liquidity game to UK government bond market makers.  We extend the analysis to the Bayesian setting, where players are of two types: large banks and small banks.  We are able to determine basic features of the new game paradigm, which align with current regulatory frameworks. Section 5 concludes with implications and suggestions for further work.


\section{Related Work on Games and Bond Markets}
Games of chance and strategy, game theory, was formally introduced by von Neumann and Morgenstern (1944) \cite{von1947theory}. Work focusing on the strategic interaction of rational players remains the foundation for the systematic study of strategic interactions in various fields, including economics, political science, and biology.  In the study of rational behaviour of players, game theory concerns itself most frequently with games characterized by exogenous utility functions. In these games, players’ preferences and payoffs are predetermined and external to the game's structure.  In keeping with traditional view points of strategic games \cite{fudenberg1991game}, each player aims to maximize their own utility, conditional upon the strategies employed by others.  Utility functions can be of the form of preference, value, price or collective co-operation explicitly for individual utility. Typically these game theoretic models rely upon assumptions of complete information (especially concerning the preferences of other players) and perfectly rational players.  These specific features tend to make games tractable but limit their applicability to real-world scenarios where uncertainty, limited (or “bounded” rationality”) and preference that evolve with time through the game.  Our contribution is to define related concepts to a real-world problem.

\subsection*{Exogenous vs Endogenous utility function}
Recent literature in the area of dynamic utility functions has delved into the dynamic nature of players preferences and the affect of environments in which players interact.  This has led to the field of research into models featuring endogenous utility functions, where players’ utilities are conditional or otherwise affected by the game's progression and the strategies employed by all participants i.e. utility functions are influences by the actions taken within the game itself. 

Adaptive preferences are one such example (\cite{AdaptiveGame}, \cite{boggess2022toward}, \cite{suryanarayana2022explainability} ). \( U_i(x, S) = ax + bS \). Here, \(U_i\) is the utility of player \(i\), \(x\) is the player's payoff, and \(S\) is a strategy profile of all players. The utility function is influenced by the chosen strategies within the game.

Likewise, the general and growing class of learning models; whether through reinforcement learning,  counter factual regret minimisation or other such method, players learn over time, and their utility functions evolve based on the history of play or observed outcomes. \cite{vadori2022multiagent}, \(   U_i(x, H) = ax + bf(H) \). Here, \(H\) represents the history of play, and \(f(H)\) is a function that captures how the history influences the player’s utility.

Finally, evolutionary game theory extensively utilised in applications of biological research.  Here the utility function represented a “fitness” of a player \cite{sandholm2020evolutionary} and that is conditional on the population share of different strategies, where strategies evolve over time in line with their success or failure (a “survival of the fit est” concept). Generally applicable to very large populations of players (i.e. species or humans) formally utility functions here are of the form \(  U_i(x, P) = ax + h(P)  \).  Here, \(P\) represents the population shares of different strategies, and \(h(P)\) is a function capturing the impact of these population shares on the player’s utility.

Common to all methods is that players are no longer passive recipients of fixed payoffs; instead, players actively shape their payoffs and these payoffs are intern shaped by the unfolding game dynamics. Our contribution to this work is to introduce an additional endogenously characterised utility function of a liquidity game, where by the utility function is the initial holding of players and characterised directly by the maximum action value they play within the game.   The game is non-cooperative reflecting the inherent tension and competition between players and the staunch opposition of such cooperation by financial regulatory bodies and governments.

\subsection*{Strategic and Extensive form games}
A short discussion between the work on strategic and extensive form games is warranted.  Historically strategic games are thought to have a fixed strategy, at inception of the game \cite{Shoham_2008}, where as extensive form games, in a reductionist manor, be considered games where players make new decisions each playing of the game \cite{ritzberger2016theory}.  Our contribution to the literature is the inclusion of a new formalism of a game, one which by definition falls into the “strategic” categorisation though with the benefit of algorithmic analysis can be easily extended to extensive form in order to check for boundedness of strategy space. 

\subsection*{Game theory and Auctions}
Much has been written recently about auctions. Within UK government bond markets no auction process takes place between market makers. However, it is an often discussed topic and warrants a brief discussion.  Recent work by \cite{DBLP:journals/corr/ChenGL14} highlights how much work on auction mechanisms contain strong assumptions, such as the truthfulness of agents and reduced form dynamics (e.g. selling identical items to uniform bidders).  The fact that many auctions seek to find fair or optimal ways to distribute fixed number of goods over a fixed number of bidders from a constrained or singular seller, means the formulation of this field does not lend itself naturally to bond markets.  Additional work by \cite{Hsu_Huang}, \cite{10.1145/2600057.2602883}, covers various assumptions and such simplifications of specific auctions.  The work into prior-free auction mechanisms also contains similar assumptions proving not helpful for bond market modeling (\cite{Sayan}).

Research into the sorts of information buyers can reveal in order to illicit a better outcome from sellers is growing in response to European regulation into e-commerce platforms targeting individuals \cite{Halpern2021CanBR}, \cite{Cai_2020}.  Again, however, research focuses on some form of problem reduction such as small scale market interactions or single product sellers etc. The notion that a buyer in an auction can voluntarily disclose information (for whatever reason) begs the question if personalised pricing of products is possible \cite{Lee_2019} with work specifically to matching platforms \cite{Ali_2019}. Various works look at the impact of incomplete information or revealed information and how a single seller may price discriminate.   Recent work by \cite{ijcai2023p293} discussed  where markets could evolve to if auction processes were ever sort to be applied widely to secondary trading of bond markets in the UK. Likewise \cite{ijcai2023p296} and \cite{ijcai2023p312} discusses features of two player auctions with varying degrees of game theory application.  Additionally, work looking into trading networks and game theory \cite{RePEc:ucp:jpolec:doi:10.1086/673402} and \cite{ijcai2023p319},  look to model similar trading concepts but often with feature sets and player dynamics distinct to our own work. 
Our contribution is rather than advancing auction methods, to re-examine the game set up and define a new form of utility mechanism based on existing real world applications. 

\subsection*{A brief review of the UK government bond market}
It is not necessary for the reader to be familiar with financial markets or bonds in particular.  However, it is worth discussing briefly the design features of the markets in which bonds are transacted.  Bond markets exist primarily in two forms: open electronic trading platforms resembling stock exchanges, and over-the-counter (OTC) markets based on bilateral negotiations. We focus on the latter.  
UK government bonds, referred to as "gilt-edged" securities, trace their origins to physical bond certificates representing borrowing by the UK government, and with gold (gilt) edges as a mark of authenticity. The UK government division, HM Treasury, through its Debt Management Office (DMO), issues and manages these bonds, known colloquially as gilts. The DMO sells gilts via its network of 17 GEMMs, regulated by the Financial Conduct Authority and the Prudential Regulation Authority (PRA), itself a subsidiary of the Bank of England overseeing financial market infrastructure. 

As licensed institutions, GEMMs have a fiduciary duty to continuously provide trade quotes to bondholders, and all client trades must be inter-mediated by a GEMM. However, GEMMs have limited capacity to accumulate large bond positions.  Whilst beyond the scope of this paper, the reader should appreciate that such limitations on bond holdings are themselves governed by legal requirements and banking regulators - all of whom must be complied with.  As such, GEMM's need to balance their portfolios, often in in a relatively short amount of time, meaning they are forced to engage in inter dealer trades with other GEMMs.  They do this in an anonymous fashion to conceal strategic information, a service which is provided by inter dealer brokers.  No GEMM can be an inter dealer broker, and inter deal brokers are separately regulated by the Bank of England. Our contribution is to model the strategic decisions processes that GEMM's make when trying to balance their portfolios through this anonymous mechanism.  

The nature of this complex and opaque market place, means that it is not possible to directly observe or model the interactions of agents within the paradigm.  Our contribution to analysing this dynamic is to propose a theoretical framework based on game theory to describe stylised facts and visible market legal requirements.  

\subsection*{Mechanism Design and Bond markets}
Our work touches on several strands of the literature. First, our motivation for researching this intricate and systematically important market stems from recent papers (\cite{BOE}, \cite{BOEOpps}, \cite{Bessembinder2020}, \cite{Biais2019}, \cite{Levine2022}). In depth analysis of related markets and the recent Gilt crisis of 2022 is covered extensively in \cite{Pinter2023}, highlighting the comparative lack of detail in this area.  Additional work focused on the US markets and their similar crisis of liquidity provision by market makers is well covered by seminal research by Darrell Duffie, most recently in \cite{Duffie2020}.

Secondly, the innovative use of game theory can be extended to agent based modeling to replicate and understand financial markets, thus expanding work included in (~\cite{Axtell2022},~\cite{Collver2017},\cite{Oriol2019}), and capital flows \cite{Braun-Munzinger2016}
\cite{Kirilenko2018}.  Whilst not covered here, the use of ABM's has been explored by the authors at length for its helpful practical elucidation of dynamic behaviours.  Our contribution here rests of the analysis of market crisis, using concepts of agents and game theory which extend
\cite{Paulin2019a}.  

Specifically, thirdly, within the realm of bond market trading, the application of game theory has garnered significant attention and has led to a range of analytical models. Our approach markedly differs from the models proposed by \cite{Carfi2012} and \cite{CarfiB2012}, who apply game theory to government bond markets. Their analysis narrows its focus on the price stabilization dynamics among three pivotal agents: two market makers and an abstract entity termed "economy in crisis." Instead of including bond buyers and sellers, they conceptualize trading as a trilateral interaction, with the crisis playing the third role. This highlights a specific contribution of our work; we specifically focus on the analysis of liquidity games, rather than games of price discovery or trading. In our contribution utility functions of plays are based on the need to rid themselves of excess assets, in order that they might go back to their primary focus, namely facilitating and trading with clients.

Our contribution to the literature is therefore to introduce the concept of a Liquidity Game (inspired by real life markets) and to document the potential that this new game approach can bring to better analyse financial markets of national importance.



\section{Method: Rules of the Game}
We assume that players have the freedom to play \textbf{heterogeneous strategies}. The game proceeds based on players strategically selecting a subset of their asset to play in the game (in this case, a parcels of bond). There is no price for the asset played. The only thing exchanged between players are the asset (bonds) themselves.  All players aim to trade as quickly as possible and to reduce their balance \(B_i\).  Agents do not wish to reveal their entire holdings to other market participants to maintain an information asymmetry advantage and meet regulatory requirements. Each player aims to reduce their holding as close to zero as possible, without any ever crossing over zero threshold. Players make plays based on assumptions regarding other traders' holdings to achieve specific objectives without revealing one's complete position. That is, players select an \(a_i\), a bond parcel constrained by their own amount of holdings.  All agents are strictly rational, and will only act if the parcel of bonds offered by the other player also improves their welfare. This strategy is known to all and players can either accept the parcel of bonds offered to them, or decline. They cannot improve their welfare with any other mechanism.

\section{Game Equilibria}

\subsection{Perfect Information Game}
If in financial markets,  "Liquidity is confidence" \footnote{https://www.federalreserve.gov/newsevents/speech/warsh20070305a.htm}, 
then solution concepts and equilibria of our game is the manifestation of financial stability. 
We show that when relaxing the concept of imperfect information, our bilateral game, of two players with equal and offsetting positions, can be shown to have a pure Nash equilibrium when \(B_i\) = 2 and \(B_j\) = -2.  Likewise, a mixed strategy Nash equilibrium is found when \(B_i\) = 3  and \(B_j\) = -3.  Whilst trivial to prove, it is illustrative of the game dynamic that the following payoff matrix is achieved:


\textbf{Payoff:} \textit{Recall the rules of game - one player has a positive holding and seeks to reduce it to zero. The opposing player has a negative holding and seeks to reduce it to zero. All bond play actions are in absolute terms.  If player 1 plays a absolute quantity less than that of player 2, the quantity of player 1 is the successful play under the rules of welfare improvement and hence the payoff is equal to the absolute quantity (as player 1 satisfies their utility function and so does player 2).  If player 1 plays a quantity greater than player 2's, no trade can occurs as player 1's quantity is larger than that of player 2 and hence player 2 can not accept any of the quantity.}
\[
\begin{array}{c|cc}
    & \textbf{Player \(N_j\)} &  \\
    & \text{ \(a_j\) = -2} & \text{ \(a_j\) = -1 } \\ \hline

\textbf{Player \(N_i\)} & & \\
 
\text{ \(a_i\) = 2} & ( 2, 2) & (0,0) \\ \hline
\text{ \(a_i\) = 1 } & ( 1, 1) & ( 1 , 1) \\
\end{array}
\]

\textbf{Equilibrium}
From the above matrix we can see that there is a pure Nash equilibrium of \(a_i, a_j\) where both are equal to (2,2) and also where \(a_i = a_j = 1\). That is somewhat trivial and intuitive - conditional on player 1 knowing with certainty that they have a holding of 2 units, and player 2 has a holding also of negative 2 units - their optimal strategy is to play 2 units. We can see that for player \(N_j\) playing \( |a_j| = 2\) is at least as good as \(|a_j| = 1\), and sometimes better, depending on what player \(N_i\) chooses. By analysing the mixed strategy equilibrium for the players (see proofs contained in supplementary material) we can see that the probability of \(N_i\) playing "2" is \(0\) and for \(N_j\) the probability is \( \frac{1}{2} \). 


An extended version of the above game where \(B_i\) = 3  and \(B_j\) = -3 (see supplementary material for data and proof), using a method of elimination, we identify elements on the diagonal of the matrix ( i.e. (3,3), (2,2) and (1,1) - noted with an asterisk ) to be pure Nash equilibrium. A mixed strategy equilibrium is found with the following weights {\(q_3 =\frac{1}{3}\), \(q_2=\frac{1}{6}\) and \(q_1=\frac{1}{2}\)} and {\(p_3 = 0\), \(p_2 = 0\) and \(p_1 = 1\)}.  In all cases, the probability increases with decreasing size of plays. 


\subsection*{Consequences and interpretation}
As is so often the case, equalibria can exist which do not necessarily produce the highest payoff for each player. It is also evident that for player 1, it is at least as advantageous for them to play a holding less than their full desired amount. The maximum payoff is achieved in both cases when player 1 is confident in playing their highest bond holdings - something that can only result in a positive payoff if they know that the other player is doing the same thing. In a real world context, this sort of dynamic would tend to suggest that players could profitably gain from knowing what the other player intends to do. Within typical regulated financial markets this sort of knowledge is not lawfully able to be used, producing a situation where we see that players are incentivised to seek out such non-lawful information.

\subsection{Bayesian framework: What type of bank is my opponent ?}

In the more realistic scenario, player 1 does not have complete information about the other but has some idea or information about the type of player they might be. Within financial markets, both in practise and in regulatory frameworks, distinctions are drawn between market participant types. The most common legal distinctions are between participants of differing "sophistication". Agent sophistication is bifurcated into "Eligible Counter parties" or "Professional Clients" as defined in MiFID II directive and implemented by the UK Financial Conduct Authority (FCA) \cite{tpicap2021best}. 
 Under FCA rules (COB 3.6.2) \cite{fcacobs2024} any GEMM will meet the definition of an "eligible counter party".  These designations seek to provide differing legal protections.  Additionally, within markets for eligible counter parties it is common for market participants to have prior information and expectations about the trading size of a counter party.  We explore this possibility and a game theoretic method further in the next section. 


\subsection*{How big is my opposing player ?}

As we can see above, an important factor on the game is the size of \(a_j\). We can see from above that should we find a situation where, for example, \(|B_i| = 10*|B_j|\), then any resulting action set \(A\) with \(a_j \in {1 \dots |B_j|}\) will have a cardinality of \(\frac{1}{10}\) of the elements \(a_i \in {1 \dots |B_i|}\).  Given that trades can only occur when players play a strategy such that \(|a_i| \cap |a_j|\) and that the quantum of \(a_n\) is determined by the size of GEMM, it makes sense to consider GEMM's of types based on a size distinction. We thus model a game where player 1 can be of two possible types: a large systemic bank with a typical and well established large trading size or a small bank with a significantly smaller typical trading size. Depending on this fact, player 2 will respond conditional on their expectation about the size of player 1. First let us explore this theoretically, before discussing the knowledge about player size which is public information for real world GEMMS. We analyse the Bayesian payoff functions below where we consider two groups.  We assume that there is a "large" group, and a "small" bank group. Our payoff functions are assumed based on the concept that should one player make a bond play that is larger than the other player, no trade can occur by game rules and thus no utility is derived. Note: for the purposes of introducing this game, we do not consider any penalty function that might arise from the act of \textit{trying to play}.  In extensions of our work beyond this submission, we envisage addressing this further.

In our current framework, we assume the following payoffs based on two types of player: Large banks and small banks. The former, we term type (a) and it is characterised by larger holdings.  The later we style as type (b), characterised by smaller balances.  The strategy that players can thus choose generalises to a notion of playing a "high proportion" of their holding or playing a "low proportion" of their holding. This analogises the trivial example above where players each have just 2 (or 3) bonds, and they can choose a strategy of playing a high amount (perhaps 3 in the trivial case) against a "low" amount of just 1.  We remind the reader of the asymmetry within the rules of the game that only allow "all or nothing" acceptance of a play.   
\\

\textbf{a type: \(p\) probability that player 2 is a LARGE bank}
\[
\begin{array}{c|cc}
    & \textbf{Player \(N_j\) = Player2} &  \\
    & \text{ \(a_j\) = high} & \text{ \(a_j\) = low } \\ \hline

\textbf{Player \(N_i\)} & & \\
 
\text{ \(a_i\) = high} & ( 10, 10) & (0,0) \\ \hline
\text{ \(a_i\) = low } & ( 6, 6 ) & ( 5 , 5) \\
\end{array}
\]

\textbf{b type: (1-p) probability that player 2 is a small bank}
\[
\begin{array}{c|cc}
    & \textbf{Player \(N_j\) = Player2} &  \\
    & \text{ \(a_j\) = high} & \text{ \(a_j\) = low } \\ \hline

\textbf{Player \(N_i\)} & & \\
 
\text{ \(a_i\) = high} & ( 0, 0) & (0,0) \\ \hline
\text{ \(a_i\) = low } & ( 5, 4) & ( 0 , 0) \\
\end{array}
\]

There is uncertainty over what type is player 2, but we can observe from the above payoffs, that player 2 knows what type they are and can see that in either case playing "high" is the weakly dominant strategy for them.  Looking for a BNE we can see that the probability that player 2 is of the large group, a, is \(p = \frac{5}{9}\), and small trading size, b, is \((1-p) = \frac{4}{5}\). 


\textbf{Therefore in summary the BNE is:}
\begin{itemize}
    \item If player 2 is type "a" (large), they strategically choose to play "High"
    \item If player 2 is type "b" (small), they strategically choose to play "High"
    \item Player 1 will chose "High" if the chance that player 2 is from the large group is \( p > \frac{5}{9}\), "low" if \( p < \frac{5}{9}\) and a equal mix if the chance is exactly \(p=\frac{5}{9}\)
\end{itemize}

From this analysis we can infer that player 1 has a preference in terms of payoffs generally for player 2 to be a type "a" player (large). Player 2's preference for strategies is to play a game of "high" plays but for player 1 their preference is determined by whom they are playing against (trading with). This demonstrates a well document preoccupation in finance for knowing the size, or type, of opposing party to a trade in a bilateral market place. See \cite{MASSA200399}

\subsection*{Real world scenario: GEMM traders}
A feature of the UK government bond market is the lack of available trading data, in part due to the historical design of the market.  The bilateral, non-exchange based mechanism we seek to model here, in reality produces sparse and non contemporaneous data points. The only source of "truth" in trading data is that available only to the Bank of England. Extensive research conducted by the Bank of England extensively discusses this drawback in \cite{Pinter2023}. 

However, the list of GEMM's in the UK is publicly available \footnote{https://www.dmo.gov.uk/responsibilities/ \\ 
gilt-market/market-participants/)} and comprises 17 banks, with 6 that would typically be considered the larger sized trading firms (Citi, DeutschBank, Barclays, JPMorgan, UBS and Lloyds Bank). Making use of stylised facts, we see that close to 35\% of all GEMMs are type "a" (large) and 65\% would be considered smaller, type "b" firms. Using these approximately we look again at a BNE analysis:

\textbf{Note:} For brevity we set "a" type p = 35\% (LARGE player 2 - denoted "L"), "b" type with 65\% probability (small player 2 - denoted "s"). Results are rounded to 1 decimal place. In reference to strategy, "High" refers to playing a high proportion of their balance, and "low" likewise refers to a strategy of playing a low amount of the players balance. All game rules remain the same.

\[
\begin{array}{c|cc|ccc|ccc}
    & \textbf{Player \(N_j\)} &  \\
    & \text{L+High} & \text{L+low}   & \text{s+high} & \text{s+low}  \\ 
    \hline

\textbf{Player \(N_i\)} & & \\
 
\text{ \(a_i\) = high} & ( 3.5,3.5) & (3.5,3.5) & (0,0) & (0,0)   \\ \hline
\text{ \(a_i\) = low } & (5.4,5.4) & (2.1,2.1)& (5, 4.4) & (1.8, 1.8)  \\
\end{array}
\]

\subsection*{Payoff analysis}

For player 2 the picture is more nuanced:  they know player 1 is a large trading counterpart (in our example that is), but also understands that player 1 is aware of the market imbalance between the number of large and small players and will probably assume that player 2 is a small player (due to market probability). Working with that assumption, player 2 can reasonably infer that player 1 will play a low amount (relative to player 1's most probable type). In that case, regardless of what actual true size player 2 is, it would be best they adopt the strategy of playing a high amount of their true quantity.  These results are reflected in the payoff matrix above.

However, it is interesting to see that the combination of situations that leads to the highest BNE payoff is where player 1 is a large player but plays the "low" strategy and where player 2, regardless of their categorisation, will always do better playing "high" strategy.  If player 2 is large, then there is no risk to this as player 1 is large and therefore compatible.  If player 2 however is small then they are also safe playing the higher quantity.

From this we see that in general we can suggest that player 1, being of the large type, always plays "low" and player 2, regardless of their own type, always play "high" to maximise their pay offs.  However, playing "low" for player 1 is not a strictly dominant strategy due to the situation where player 2 might be a large player and be interested in playing low also.  

Clearly here we have not defined some reference of what constitutes "high" and "low" and what is small and large. Crucially we have not clearly defined what bond quantity constitutes a high or low amount for each large and small player. Instead we have relied upon a fuzzy notion of equivalence to illustrate the potential.  Further refinement will likely be possible with iterative computational analysis. In practise, these group assignments are frequently imprecise and thus may not be so far from practical reality after all. 

One further solution concept to consider is borrowed from quantitative finance. The concept of "hit ratio"; utilizing the probability weighted frequency of "success" as a measure of trading strategy \cite{DICHTL201688}.  In game theoretic terms we interpret this to be what percentage of our payoff functions are non zero. In this case we see overall the bilateral interaction has 75\% of interactions resulting in non zero payoffs. While potentially a novel application from a game theory perspective, this forms the basis of an argument for stability.  We consider where this example deviates from reality: in addition to the above mentioned concerns regarding the fuzzyness of boundary assignments of group types, we also come across the issue that player 2 does not actually know the identity of player 1 or their type.  This is a more fundamental issue for our modeling and will be addressed next.

\subsection{Market design questions: Incomplete information and uncertainty about both player groups}
Whilst it could be concluded, that any player in a game must certainly know which type they belong to, when games are externally observed it is a natural extension to analyse the situation where both players are of uncertain type. This is indeed the situation that regulators or market designers find themselves in. Ideally, market design should align with what any player would prefer to do, a real world implication of trying to find a equilibrium.  Utilising the market structure (large GEMM's = 35\% , small ones 65\%), multiplying the above matrix through we have the following (note: rounding to 1 decimal place for readability); \\
\\

\resizebox{.91\linewidth}{!}{$
            \displaystyle
\begin{array}{c|cc|ccc|ccc}
    & \textbf{Player \(N_j\)} &  \\
    & \text{L+H} & \text{L+l}   & \text{s+H} & \text{s+l}  \\ \hline

\textbf{Player \(N_i\)} & & \\
 
\text{L+H} & ( 1.2, 1.2)  & (1.2, 1.2) & (0,0) & (0,0)               \\ 
\text{L+l} & (1.9, 1.6) & (0.7, 0.7) & (1.8, 1.5) & (0.6, 0.6)    \\ \hline

\text{s+H} & (2.3, 2.3) & ( 2.3, 2.3) & (0, 0) & (0,0)      \\
\text{s+l} & (3.5, 3.1) & (1.4, 1.4) & (3.3, 2.8) & (1.1, 1.1)  \\
\end{array}           
         $}
\\
\\

\textbf{Solution concept and Summary}
Recall that the payoff matrix is populated by the utility of each player, determined endogenously by the size of the play, or action \(a_i\) successfully carried out.  A market regulator might be most interested to understand how the highest payoffs could thus be generated, give they represent the actions with the largest volumes of trades (again, \(a_i\)) and this in tern represents the highest turn over possible. As such, a scenario where both players are large players (the upper left quadrant) does not contain all of the highest payoffs.  In fact, summing over all volumes transacted (again relying on payoff to measure volume of bond traded), we see that in the scenario with only large players a total of 9.7 units would transact only slightly more than the worst overall combination of both small players in the bottom right quadrant with a total of 8.3 units.  Given that the total summed payoff for the whole system is 41.1, we see that from a standpoint of maximising volumes, a regulator would want some mixture of large and small players.  Or, more strongly put, markets appear more stable when large players are not allowed to dominate trading, despite how counter intuitive this is to the goal of maximising transfer of bonds in totum.  

\subsubsection{Player size}
Contrast the "large only" quadrant to the quadrant with the highest total, the bottom left, where player 1 is from the majority "small" bank group and player 2 from the minority "large" group. The asymmetry in payoffs is a direct feature of the game rule regarding the second player being only able to accept a bond parcel up to and including their total need. (contained also in the rule regarding no player balance may cross a zero threshold from one sign to another). In practical terms, this rule prohibits a player, market maker, from clearing a balance to only then transform their holding into a balance of the opposite sign, effectively entering into a second balance.  Concluding as we can that from a design point a mixture of trading types appears desirable, including the "small" types.  Whilst outside the scope of this paper, in financial terms some of the smallest players, high frequency trading firms, trading in the smallest permissible unit in markets, are often derided as providing negative value \cite{Kirilenko_2017}, \cite{Kirilenko_2017} and \cite{Kirilenko_2018}. Our initial analysis would seek to challenge the broad assumption that "bigger is better" and that a mixture of trading size is preferable.

\subsubsection{High or Low trading proportion strategy}
Turning to the strategy of trading either a high or low proportion of ones bond holding, both players choosing either both "high" or both "low", results in the lowest combined volume. Thus it can be inferred that some mixture of participant strategy is desirable.

We see here, as in the simpler case, player 1 \(N_i\) choosing a playing strategy of being in the "low" trading size (regardless of player size) will dominate the results of being in the higher trading proportion group in all but one specific scenario: that in which they are trading with a player of the same size and strategy .  

In this way we see that in some ways it is preferable for player 1, the initiating player in the simultaneous game, to to be a smaller player.  Within our frame work, like the games that inspire it, agents cannot choose their type designation. Furthermore, to do so (through possible strategy selection) would allow a degree of possible game manipulation. This would be prohibited in financial markets but also has been shown to affect reputations (recall players learn the identity of others after a successfully play, leading to a degree of information leakage). As shown in \cite{MASSA200399}, this is known to be detrimental to players. 


We can see from the Bayesian game above that players who are small do better than larger players. In addition, player 2 is better playing a "high" strategy regardless.  Player 1 however is better under all scenarios to play a "low" strategy.  This in some way is not surprising given the game set up - those precise combinations increase the chances that the respective players' action spaces intersect, where an intersection leads to a trade and therefore a quicker removal from the game.  Further more, the most successful quadrant is that where the initiating player is a small bank, and they engage with a large bank. Given the proportion of the GEMM market composition, this would seem to substantiate the current market composition as determined by the Bank of England.

\section{Further work}
Our initial analysis of solution concepts suggests further work to evaluate and describe the mechanism of the game.  We see natural extensions to Stackelberg game variants and the use of computation game theory to analyse if learning polices is possible in repeated games.  Additionally, early work suggests the application of agent based models to look to better understand the nuanced interactions between players and to extend the model to \(N_i \geq 2\) players. 

\section{Conclusion}
We introduce a new type of game, the Liquidity Game, inspired by real-life opaque bilateral bond market trading activities between UK market makers.  The goal of this paper was to introduce the readers to a new formalism of a strategic game with endogenous utility functions and to describe an AI technique for a novel application domain.  Our technical contribution rests on a new game formalism, generalised to strategic games of incomplete information, price-free asset (or goods) transfer.  We propose a Bayesian equilibrium as a solution concept but begin our introduction to the reader by means of two trivial examples to illustrate the game dynamics and payoffs of each player, under perfect information, but without cooperation.  We extend this consideration to the practical real-world example of Bayesian Liquidity Games between two market makers, belonging to one of two types of player: a large bank or a small bank. Again we take inspiration from the real world design detailed by the Bank of England and the proportions of large and small participants in the market place (17 firms in total). Proof's, derivations, data, and source code are provided in supplementary materials.  
Having described the systemic importance of a well functioning market for UK government bonds, we discuss the strategic importance of bilateral mechanisms being able to clear their accumulated balances in a trusted manner, namely within a mechanism framework that gives them each a high chance of being able to trade.  We extend this game theoretic work to discuss implications for market design by regulators, looking to find inherently stable designs with the ability to support large, if not maximal, liquidity.



\bibliographystyle{named}
\bibliography{bib}
\end{document}


\maketitle

Thank you for taking the time to refer to this document. For ease of reference, a short table of contents is below. This document contains proofs, code notes, derivations and details regarding data used in, and supplementary to, the submission entitled "Non cooperative Liquidity Games and their application to bond market trading". All python code can be found in the zip file and referenced below.

\\
\\

\tableofcontents
\\
\\
\section{Code: Empirical Computational Game theory code}
In support of proofs and validation of the associated submission the following is a set of attached files with descriptions:

\subsection{Nash and mixed Nash computational analysis}
File: \textbf{IJCAI Mixed Nash 2 player.py}

\subsection{Bayesian Nash Simulation and Analysis}
See python code File: \textbf{Bayesian Nash simulation.py} - this file contains the simulator of the main game with a range of values for each player.  File: \textbf{"Analysis Bayesian Nash simulation.py"} provides analysis code and output is demonstrated in file \textbf{"example BN output.txt"}. Also, a sample of a Bayesian setting with 10,000 simulations and initial \(B_i \in {1 \ldot 1000\) and \(B_j \in {-1000 \ldot -1\) can be found in file \textbf{"Sample Bayesian simulation 10k.xlsx"}.

\section{Derivation of NE for \(B_i\)= 2 and \(B_j\)= -2 }
This derivation relates to details in \textbf{4.1: Perfect information games} of the main paper (section beginning line 362 and following). 

\subsection*{Trivial example: Mixed Strategy Equilibrium}
\begin{proof} Mixed Strategy Equilibrium for player \(N_i\)

    \( E(U_B_{j_{2}}) \) = \( E(U_B_{j_{1}) \)  \\
    \( E(U_B_{j_{2}}) \) = \( f(\theta_{i_2}) \) = \(\theta_{i_2} * 2 + (1 - \theta_{i_2}) *1 \\
    \( E(U_B_{j_{1}}) =  f(\theta_{i_2})\) = \(\theta_{i_2} * 0 + (1 - \theta_{i_2}) *1 \\
    \textit{therefore}  \( \theta_N_{i_2} = 0\)  \\
    
    \textit{Similarly for \(N_j\):} \\
    \( E(U_B_{i_{2}}) \) = \( E(U_B_{i_{1}}) \)  \\
    \( E(U_B_{i_{2}}) \) = \( f(\theta_{i_2}) \) = \(\theta_{i_2} * 2 + (1 - \theta_{i_2}) *0 \\
    \( E(U_B_{i_{1}}) =  f(\theta_{i_2})\) = \(\theta_{i_2} * 1 + (1 - \theta_{i_2}) *1 \\

    \textit{therefore}  \( \theta_N_{j_2} = \frac{1}{2}\)  

\label{MSE2}
\end{proof}
The above details the derivation of values given on line 395 of the main paper.

\section{Derivation for solution and payoff for \(B_i\) = 3  and \(B_j\) = -3}

\[
\begin{array}{c |cc|ccc}
    & \textbf{Player \(N_j\)} &  \\
    & \text{ \(a_j\) = -3} & \text{ \(a_j\) = -2}  & \text{ \(a_j\) =- 1 } \\  \hline

\textbf{Player \(N_i\)} & & \\
 
\text{ \(a_i\) = 3} & ( 3*, 3*) & (0,0) & (0,0) \\ \hline
\text{ \(a_i\) = 2 } & ( 2, 2) & ( 2* , 2*) & (0,0) \\
\text{ \(a_i\) = 1 } & ( 1, 1) & ( 1 , 1) & (1*,1*) \\
\end{array}

\label{PNE3}
\]

Assigning probability \(q_3\) to playing \(a_j = 3\) and \(q_2\) to playing \(a_j = 2\), likewise probability \(p_3\) to \(a_i=3 \) and so forth we find that:

\begin{proof}{Mixed Strategy Equilibrium for player \(N_j\)} \\
\( E(U_B_{i_{3}}) \) = \( E(U_B_{i_{2}) \) = \( E(U_B_{i_{1}) \)  \\
\( E(U_B_{i_{3}}) = 3*q_3 + 0*q_2 + 0*(1 - q_3 - q_2) \) \\
\( E(U_B_{i_{2}}) = 2*q_3 + 2*q_2 + 0*(1 - q_3 - q_2) \) \\
\( E(U_B_{i_{1}}) = 1*q_3 + 1*q_2 + 1*(1 - q_3 - q_2) \) \\
\textit{therefore}  \( q_3 = \frac{1}{3} \) , \(q_2 = \frac{1}{6}\) and \(q_1 = \frac{1}{2}\) 
\end{proof}

\begin{proof}{Mixed Strategy Equilibrium for player \(N_i\)} \\
\( E(U_B_{j_{3}}) \) = \( E(U_B_{j_{2}) \) = \( E(U_B_{j_{1}) \)  \\
\( E(U_B_{j_{3}}) = 3*p_3 + 2*p_2 + 1*(1 - p_3 - p_2) \) \\
\( E(U_B_{j_{2}}) = 0*p_3 + 2*p_2 + 1*(1 - p_3 - p_2) \) \\
\( E(U_B_{j_{1}}) = 0*p_3 + 0*p_2 + 1*(1 - p_3 - p_2) \) \\
\textit{therefore}  \( p_3 = 0 \) , \(p_2 = 0\) and \(p_1 = 1\) 
\end{proof}

The above derivation supports values provided in line 400 of the main paper.

\section{Simple linear programming code example for 2 player game \textbf{generalised}}

\begin{lstlisting}
    # Define Player 2's negative assets (A) and Player 1's positive assets (B)
A = 10  # Replace 10 with the actual negative asset value of Player 2
B = 20  # Replace 20 with the actual positive asset value of Player 1

# Define the coefficients of the objective function
# To minimize B - x, we minimize -x (equivalent to maximizing x)
c = [-1]  # Minimize -x

# Define the inequality constraints
# x - A <= 0 (rewritten as -x + A >= 0 for linprog) and x <= B
A_ub = [[1], [1]]  # Coefficients for the inequalities
b_ub = [A, B]       # Upper bounds for the inequalities

# Define the bounds for x
# x should be non-negative
x_bounds = (0, None)

# Solve the LP
result = linprog(c, A_ub=A_ub, b_ub=b_ub, bounds=[x_bounds], method='highs')

# Output the result
if result.success:
    transfer_amount = result.x[0]
    print("The amount Player 1 should transfer to Player 2 is:", transfer_amount)
else:
    print("No feasible solution found.")

\end{lstlisting}
